\documentclass[aps,pra,onecolumn]{revtex4}
\usepackage{amsmath}
\usepackage{amssymb}
\usepackage{graphicx}
\usepackage{ulem}
\usepackage{color}
\usepackage[dvipsnames]{xcolor}
\definecolor{orange}{rgb}{1,0.5,0}

\begin{document}
\setcounter{secnumdepth}{-1} 

\title{Synchronization, quantum correlations and entanglement in oscillator networks}

\author{Gonzalo Manzano}
\author{Fernando Galve}
\author{Gian Luca Giorgi}
\author{Emilio Hern\'{a}ndez-Garc\'{\i}a}
\author{Roberta Zambrini*}
\affiliation{Institute for Cross Disciplinary Physics and Complex Systems,
IFISC (CSIC-UIB), Palma de Mallorca, Spain}

\begin{abstract}
Synchronization is one of the paradigmatic phenomena in the
study of complex systems. It has been explored theoretically and
experimentally mostly to understand natural phenomena, but also
in view of technological applications. Although several mechanisms
and conditions for synchronous behavior in spatially extended
systems and networks have been identified, the emergence of this
phenomenon has been largely unexplored in quantum systems
until very recently. Here we discuss synchronization in quantum
networks of different harmonic oscillators relaxing towards a stationary state, being essential the form of dissipation. By local
tuning of one of the oscillators, we establish the conditions for
synchronous dynamics, in
the whole network or in a motif. Beyond the classical regime we show that 
synchronization between (even unlinked) nodes witnesses
the presence of quantum correlations and entanglement. Furthermore, synchronization and entanglement can be induced between
two different oscillators if properly linked to a random network.
\end{abstract}

\maketitle
*Correspondence to roberta@ifisc.uib-csic.es

Synchronization is a paradigmatic and well studied
phenomenon in many biological, physical and social systems
\cite{Strogatz,Pikovsky,Arenas}, also proposed as a tool in
view of applications \cite{Athen}, but almost unexplored in the
quantum regime. Interesting exceptions \cite{entrain1,entrain2,entrain3} deal
with entrainment, i.e. synchronization as the response to an
external driver. At the microscopic level, spontaneous
synchronization was recently predicted in nanomechanical
oscillators \cite{Marquardt2011,Milburn2011} and observed
\cite{Lipson2012} in the classical regime, when quantum
fluctuations are neglected. Indeed a first approximation to a
great variety of quantum systems, such as electromagnetic modes
\cite{opt1,opt2,opt3},  trapped ions \cite{ions1,ions2} or nanoelectromechanical
resonators \cite{Plenio}, is given by a set of coupled quantum
harmonic oscillators,  susceptible to experience spontaneous
synchronization. Beyond physical systems, there is an
increasing awareness that quantum phenomena might play an
important role in terms of efficiency of biological processes.
Several examples \cite{Gauger2011,Panitchayangkoon,Engel} have shown that we may
have to consider quantum dynamics to explain biological
phenomena. 
 

A first step to characterize quantum spontaneous
synchronization, considering quantum fluctuations and
correlations beyond the classical limit, has been taken in
Ref.\cite{PRAsync} where synchronization between one pair of
damped quantum  harmonic oscillators has been reported. 
Most of the classical literature deals
with self-sustained phase oscillators modeled by Kuramoto-type
models, or with identical nonlinear oscillators studied through
the master stability formalism \cite{Arenas}. Ref.
\cite{PRAsync} takes a different view focusing on
synchronization during the relaxation dynamics of different
linear oscillators driven out of equilibrium  and exploring the
key role of dissipation.  Indeed, depending on
the 
way in which damping occurs, a pair of oscillators with
different frequencies  can show synchronous evolution emerging
after a transient, as well as robust (slowly decaying)
non-classical correlations \cite{PRAsync}. When several
dissipative quantum oscillators coupled in a network are
considered dissipation can act globally or locally (in a node)
and, depending on the correlation length in the bath with
respect to the size of the system, a variety of surprising
phenomena are observed. In this work we show how
synchronization can actually be induced by local tuning of one
(even newly attached) oscillator of a generic (regular or
random) network. Synchronous behavior emerges in the whole
network or in a part of it and  witnesses robust quantum
correlations and entanglement. Stemming both from the structure
of the system and from the form  of system-bath coupling we
further show the possibility to tune the system to
configurations in which nodes do not thermalize and relax into
a  synchronous and non-classical asymptotic state.

The form in which dissipation occurs in a spatially  extended
system 
has deep consequences.
We stress that the importance of symmetries present in the
system-bath coupling has been recognized in many contexts. In
classical systems, this fundamental issue was already discussed
in  the seminal work of  Lord Rayleigh analyzing damping
effects on normal modes in vibrating systems \cite{Rayleigh}.
Indeed, the role of dissipation to reduce detrimental effects
of vibrations is fundamental  in many areas of mechanical,
civil and aerospace engineering \cite{Engineering}. 
On the other hand, in the context of quantum systems, symmetries in the coupling
between qubits and the environment  allow for decoherence-free
subspaces \cite{dfs}, entangled states preparation \cite{Diehl,Barreiro2010}
and dissipative quantum computing \cite{cirac2009,blatt2011}. 

In the following we  show that the distribution and 
form of losses through the network amounts to 
synchronous dynamics in spite of the nodes diversity, and  robust 
quantum correlations. 
Furthermore, steady entanglement can be generated between 
unlinked nodes by properly coupling them to a network.

\section{Results}
\subsection{Dissipation mechanisms and synchronization}
We consider generic networks of $N$ nonresonant, coupled
quantum harmonic oscillators, given by the Hamiltonian
\begin{equation}\label{HS}
 H_S 
 = \frac{1}{2} \left( \bf{p}^T {\bf{p}} + {\bf{q}}^T \mathcal{H}  {\bf{q}} \right)
\end{equation}
where ${\bf{q}}^T = (q_1, ..., q_N)$ is the vector of canonical position operators and ${\bf{p}}$
are momenta, satisfying $[q_j,p_j]= i$
 (we take $\hbar=1$ throughout the paper), and
$\mathcal{H}_{m,n}=\omega_m^2 \delta_{mn} +
\lambda_{mn}(1-\delta_{mn})$ is the matrix containing the
topological properties of the network (frequencies $\omega_m$ and couplings $\lambda_{mn}$).
The eigenmodes of the system $\bf{Q}$ result from diagonalization of
this Hamiltonian through the transformation matrix $\mathcal{F}$.

\begin{figure}
\includegraphics[width=8cm]{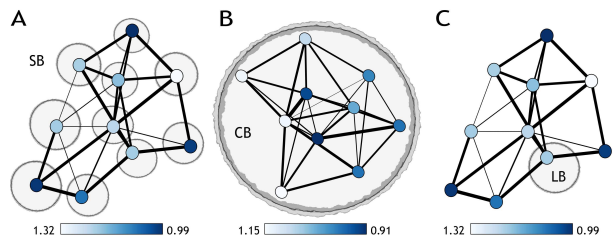}
\caption{
(A) Network of oscillators (represented by the network nodes) dissipating into separate baths (SB,
represented by the gray circles surrounding the nodes).
Links, representing couplings, have different strengths (lines thickness) and nodes have different natural frequencies (corresponding to
different colors as given in the color bar). (B) Network of oscillators dissipating into a common bath (CB). (C)
Network of oscillators with dissipation restricted to one node, local bath (LB).
\label{fig_1}}
\end{figure}


Any realistic model needs to include also  environment effects \cite{Weiss,Schloss,gardiner} and,
depending on system and bath correlations lengths, different forms of dissipation can be envisaged for
an extended network (see Methods). All units can dissipate into separate identical baths (SB),
Fig.\ref{fig_1}A, as cavity optical modes \cite{Weiss,gardiner}. Otherwise, if the coherence length of
the environment is larger than the size of the system (here given by the spatial extension occupied by
the oscillators network in Eq.\eqref{HS}), all the nodes  feel a 'similar' dissipation.  This
common bath (CB),  Fig.\ref{fig_1}B, is known to create decoherence-free subspaces \cite{dfs,Viola} 
and asymptotic entanglement 
\cite{klesse,paz-roncaglia,liu,galve2010}. A third, limiting, case (LB) in which a specific oscillator
$d$ dissipates much faster than any other node (Fig.\ref{fig_1}B) is also considered (local bath).

One of the key insights of our work comes from noting that the coupling of real oscillators to the bath (taken here to be equal, $\gamma$) differs from 
those of eigenmodes. The latter are found to be
 $\gamma\kappa_m$ (with $\kappa_m^{CB}=\sum_n\mathcal{F}_{nm}$, $\kappa_m^{SB}=1$ and $\kappa_m^{LB}=\mathcal{F}_{dm}$) meaning that, except the SB situation, the eigenmodes have
different decay rates. Then for CB and LB only the least dissipative eigenmode will survive to thermalization, thus governing the motion of all oscillators overlapping with it.  It is then
useful identifying the less dissipating normal mode with
smallest {\it effective coupling}, $\kappa_\sigma$, and also the
following one, $\kappa_\eta$, such that  $|\kappa_\sigma|\le
|\kappa_\eta|$. Further, in the case of an Ohmic bath with cut-off-frequency larger than
the frequencies of the system, the normal-mode
damping rates simplify to the form  $ \Gamma_{i} = \gamma ~
\kappa_i^2$

\subsection{Conditions for synchronization in a network}

Knowledge of the normal modes of a complex network and of their
dissipation rates (or effective couplings) allows to fully
characterize a large variety of phenomena. Indeed this is a
simple but powerful approach, even if diagonalization of the
problem needs to be performed numerically except in a few
(highly symmetric) configurations.  By the diagonalization
matrix $\mathcal{F}$ and system-bath interaction Hamiltonian we
obtain the conditions to have a dominating mode during a
transient. This mode dissipating most slowly,
$|\kappa_\sigma|<|\kappa_j|$ $\forall j\neq\sigma$, is found
either for CB or LB.  Even more important, it is possible to
identify normal modes completely protected against dissipation,
so that some network's nodes do not thermalize. Indeed, a
normal mode $\sigma$ is protected against decoherence if
$\kappa_\sigma=0$. For a pair of oscillators interacting with a
CB, this condition is accomplished only in the trivial case of
identical frequencies \cite{paz-roncaglia,galve2010} but this
is not the case when more than two nodes are considered.  We
find that asymptotic synchronized quantum states can then be
observed even in random networks where all nodes have different
natural  frequencies.

\begin{figure}
\includegraphics[width=7cm]{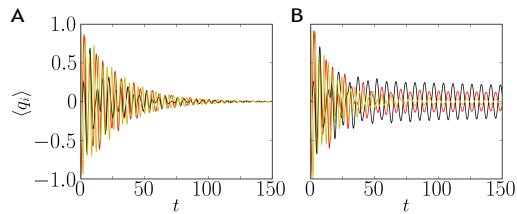}
\caption{
First order moments for initial conditions  $-\langle q_1\rangle=\langle q_3\rangle=1.0$, $\langle q_2\rangle=0.0$, and vanishing
 momenta in the case of an open chain of three oscillators with $\omega_1=1.2~\omega_2$, $\omega_3 = 1.8 ~\omega_2$,
non-vanishing couplings $\lambda_{12}=\lambda_{23}=0.4~\omega_2^2$, temperature $T=10~ \omega_2$
(Boltzmann constant taken to be unity), $\gamma= 0.07~ \omega_2^2$, bath cutoff $50~\omega_2$, for SB (A) and
for CB (B). Time is scaled in units of the inverse of the reference frequency $\omega_2$.
\label{fig_2}}
\end{figure}

Full characterization of the quantum state evolution of the
network comes from moments of all orders of the oscillator
operators ${{q}_j}$ and ${{p}_j}$  \cite{gardiner}. 
For SB, average positions (and momenta) are
characterized by irregular oscillations before
thermalization  (Fig. \ref{fig_2}A). On the other hand, for
dissipation in  CB and after a transient, regular phase locked
oscillations can arise, as shown in Fig. \ref{fig_2}B.
Synchronization  between detuned nodes can be found  during a
rather long and  slow relaxation, like in the case of
just one pair \cite{PRAsync}. Further, the oscillations can
remain robust even asymptotically if the condition
$\kappa_\sigma=0$ is satisfied.

Beyond the classical limit given by average positions and momenta, let us
now consider the full quantum dynamics stemming from the
evolution of higher moments (see SI). At the microscopic level,
quantum fluctuations also oscillate in time (even for initial
vacuum states for which first order moments vanish at any time).
This collective periodic motion is
associated to a slow energy decay and witnesses the presence of
robust quantum correlations against decoherence \cite{Schloss}.
Our approach points to a wide range of appealing possibilities
in quantum networks. In the following we show how  a whole
random network (or a part of it) can be brought to a
synchronized state retaining quantum correlations via local
tuning of just one of the nodes, or how two external
oscillators can be linked to a random network leading to their
entanglement and locked oscillations.

\subsection{Collective synchronization by tuning one oscillator}

Let us consider an Erd\"os-R\'enyi random and dissipative network
\cite{Arenas} of oscillators with different node frequencies,
links and weights (Fig. \ref{fig_1}) focusing on the
relaxation dynamics of energy and quantum correlations. The
node dynamics is mostly incoherent and even if initializing the
network in a non-classical state, quantum correlations
generally disappear due to decoherence \cite{Schloss}.
Independently on the form of the network, for dissipation in
SB, all nodes thermalize on a time scale $\gamma^{-1}$
(see Eq. \eqref{HISB} and Methods). As anticipated
before, this is not the case in the presence of a dissipation
acting not-uniformly within the network.

\subsubsection{Common dissipation bath}

\begin{figure}
\begin{center}
 \includegraphics[width=8cm]{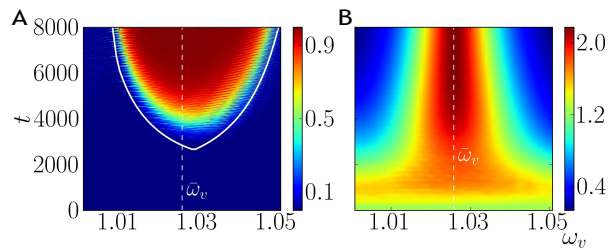}
\end{center}
\caption{(A) Time evolution of synchronization $\mathcal{S}$,
 and (B) quantum correlations quantified by the discord $\langle\delta\rangle\times 10^{3}$ , when varying one
 node frequency $\omega_v$. Results are shown for a random network (connection probability
$p=0.6$) of $10$ oscillators. Frequencies of nodes are sampled from a uniform distribution
from $0.9 \omega_0$ to $1.2\omega_0$ and couplings from a
Gaussian distribution around $-0.1\omega_0^2$ with standard deviation $0.05 \omega_0^2$. 
Environment parameters (here and in the following figures) are
$\gamma = 0.01\omega_0^2$, $T = 10 \omega_0$ and cut-off frequency $\Lambda=50 \omega_0$ (see Methods).
Time units are chosen so
that $\omega_0=1$. Collective synchronization $\mathcal{S}$ and (averaged and filtered)
discord $\delta$ (see Methods) are obtained considering all oscillator pairs of the network.
Dashed line identifies the frequency $\bar\omega_v$
for which  $\kappa_\sigma=0$. Continuous line in (A) corresponds to the estimated synchronization time
$t_{sync}$.
\label{fig_3}}
\end{figure}

In presence of CB,  an arbitrary network of $N$ nodes can reach
a synchronized state before thermalization if there is a
weakest effective coupling $\kappa_\sigma$. As a matter of fact
just by tuning one of the node frequencies, $\omega_v$, even
maintaining fixed the rest of the network frequencies
$\{\omega_{l\neq v}\}$ and its topology ($\lambda_{ij}$
couplings) it is possible to decrease the weakest coupling
$\kappa_\sigma$.   This means that an extra oscillator of
properly selected frequency $\{\omega_{l\neq v}\}$ (like a
synchronizer) can be added to a random network, even if weakly
coupled, and it will lead to a collective synchronization of
the whole system at some frequency ($\Omega_\sigma$), generally
different from $\omega_v$. 
Fig. \ref{fig_3} displays the
average global synchronization and quantum correlations
established in the network. Synchronization arises after a
transient across the whole network by tuning one of the
frequencies $\omega_v$ to a particular value $\bar\omega_v$,
while it is not present when moving a few percent away from
this value. Equivalently one could have  tuned one of the couplings
$\lambda_{vv'}$.
In the following we consider separately the case in
which $\kappa_\sigma$ is significantly smaller than the other
effective couplings and the case in which it vanishes.

Conditions for global synchronization follow from small ratio
between the damping rates of the two slowest normal modes
$R=\kappa_\sigma/\kappa_\eta\rightarrow 0$. Interestingly, this
is a necessary but not sufficient condition for $collective$
synchronization. This is due to the fact that the presence of a
slowly dissipating normal mode into the system needs to be
accompanied by a significant overlap between this mode
($Q_\sigma$), or virtual oscillator, and all the real  ones
($q_1, ...,q_N$). An analytical estimation of the
synchronization time is found taking into account both the
importance (overlap with individual oscillator) and decay of
few normal modes in the system. The time for oscillator $j$ 
to start oscillating at the less damped frequency 
$\Omega_\sigma$ reads:
\begin{equation}\label{time_sync}
 t^{(j)} \equiv \max_{\{ k \neq \sigma \}} 2 
 \frac{\log{\mathcal{F}_{j k}}-\log{\mathcal{F}_{j \sigma}}}
 {\Gamma_{k} - \Gamma_{\sigma}} \ ,
\end{equation}
maximizing over all modes $k$ different from the
slowest one $\sigma$ (see SI). Collective synchronization time
corresponds to $t_{sync} = \max_{\{j\}} \{ t^{(j)}\}$, i.e.
when even the last oscillator joins the synchronous dynamics
dominated by the less damped mode. Then phase-locking in the
evolution of all oscillators, namely in their (all order)
moments, can arise before thermalization, when there is
significant separation between largest time decays
$\Gamma_{\eta},\Gamma_{\sigma}$, and  overlap between slowest
normal modes and each system node. Global network
synchronization (see Methods) obtained from the
full dynamical evolution and the estimated synchronization time
$t_{sync}$ are in good agreement, as seen in Fig.
\ref{fig_3}A. 
For the same network, the ratio $R$, the collective synchronization $\mathcal{S}$ and mean discord $\langle\delta\rangle$ at long times
can be seen in Fig. S1 (see SI).

We now look at the quantumness of the state in presence of
collective synchronization. Generally decoherence is
independent on the specific features such as the oscillation
frequency in a system \cite{CaldeiraLeggett}. Still,
synchronization is a consequence of a reduced dissipation in
some system mode and indeed witnesses the robustness of quantum
correlations, as evident form the average discord $\langle\delta\rangle$ in
the network represented in Fig.\ref{fig_3}B. After a transient
dynamics in which the couplings in the network create quantum
correlations \cite{Plenio}, even when starting from separable states, 
discord does
actually decay to small values for $\omega_v$ different from
$\bar\omega_v$ (non-synchronized network) while it maintains
large values for the case of a properly tuned node
($\omega_v\sim\bar\omega_v$).

The case $\omega_v=\bar\omega_v$, leading to $ \kappa_\sigma =
0$, needs special attention. 
After a transient
all the nodes will oscillate  at a locked common frequency, the
one of the undamped normal mode $\Omega_\sigma$, which we call
a {\sl frozen} mode. As before (Eq.\eqref{time_sync}), the
possibility to synchronize the whole network also requires a
second condition, namely that the undamped mode involves all
the network nodes. (The case in which the latter condition
applies only to some nodes is discussed below.)
 When  both the conditions
\begin{equation}\label{eq_CB_sync}
\kappa_\sigma = \sum_{k=1}^N \mathcal{F}_{k \sigma}= 0 \text{  ,
    and  } \mathcal{F}_{k \sigma}\neq 0 \text{ } \forall k
\end{equation}
are met, there is a frozen normal mode  linking all oscillators.
This leads to collective synchronization in the whole network and
allows for mutual information and quantum correlations
remaining strong  even asymptotically, being orders of magnitude
larger than for the fully thermalized state, when
synchronization is not present (Fig.\ref{fig_3}).  The
undamped mode gives actually rise to a decoherence-free
dynamics for the whole  system of oscillators where quantum
correlations and mutual information survive.

The phenomena above are found for nodes dissipating at equal
rates into a CB, while in the presence of  $N$
independent environments (SB) all oscillators thermalize
incoherently, synchronization is not found, and decoherence
times for all oscillators are of the same order. As a final
observation 
we mention the special case in which
 the center of mass of the system is one of the
normal modes; then there will be a large decoherence-free
subspace (corresponding to the other $N-1$
 modes) but  no synchronization will appear for a CB.

\subsubsection{Local dissipation bath}

Common and separate baths correspond to two extreme situations
in which all oscillators have equivalent interactions with the
environment(s). We now consider the case of a local bath, as a
limit case in which one oscillator is dissipating stronger,
Fig.\ref{fig_1}C. 
A frozen normal mode $\sigma$ must not overlap with the
dissipative oscillator (labeled by $d$) while involving all the 
other nodes ($ \mathcal{F}_{i\sigma} \neq 0$ 
 $\forall~ i \neq d$). Then,
synchronization of the whole network (except for the
dissipative oscillator) arises. 
This occurs when
\begin{equation}\label{eq_LB_sync}
\mathcal{F}_{d \sigma} = 0 \text{ , with }   \mathcal{F}_{d j} \neq 0 
\text{   }\forall~ j \neq \sigma
\end{equation}
meaning that the undamped mode $ \sigma$ involves  a cluster of
oscillators not including the lossy one. 
We find synchronization and
robust quantum effects across the network as
for CB, with the difference that for LB the dissipating node is
now excluded (further details are discussed in SI).

\subsection{Synchronization of linear motifs}

\begin{figure}
\includegraphics[width=8cm]{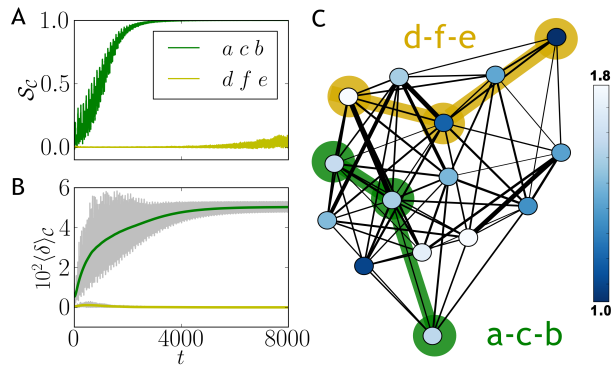}
\caption{(A) Synchronization factors $\mathcal{S}_C$ and (B) average
discord $\delta_C\times 10^2$ evaluated for linear 3-node motifs
(hence the subindex $\mathcal{C}$) in a random network (connection probability $p=0.6$) of
$15$ oscillators (shown in C). A tuned non-dissipative motif $\mathcal{C}_1$  composed by the three nodes
$(a-c-b)$ is compared with another equivalent non-tuned motif  $\mathcal{C}_2$ composed by nodes
$(d-f-e)$. Frequencies in the network are sampled from a uniform distribution from $\omega_0$ to $1.8
\omega_0$, and couplings with a Gaussian  distribution around $-0.1\omega_0^2$ with standard deviation $0.05
\omega_0^2$.  In order to avoid dissipation in the $(a-c-b)$
motif we have set $\omega_c = 1.51~ \omega_0$,  being $\lambda_{a c}=-0.09~ \omega_0^2$ and
$\lambda_{b c} = - 0.11~\omega_0^2$. Time units such that $\omega_0=1$.\label{figure4}}
\end{figure}

The possibility to synchronize a
whole network, in presence of different dissipation mechanisms,
just by tuning one local parameter opens-up the
perspective of control that can be explored considering the
dynamical variation of a control-node frequency. In particular
we find similar qualitative results both for random networks 
and for disordered lattices consisting of regular networks
with inhomogeneous frequencies and couplings, being the latter largely studied in ultracold
atomic gases  \cite{lew}. Local tuning to collective
synchronization is not only a general feature of different kind
of networks but can also be established in motifs within the
network. As we show in the following, the system can be tuned
to a partial synchronization, involving some nodes of a network
independently of the rest of it.  Indeed, even if the whole
system is coupled to a CB,  we can identify the
conditions for having a synchronized cluster, like the 3-node
linear motif in Fig. \ref{figure4}. Two non-directly linked
nodes $a$ and $b$ of the motif   are then asymptotically
synchronized through another one, here $c$,  and this leads to
a  common oscillation dynamics along the whole motif $a$-$c$-$b$. The
condition for synchronization of a cluster, namely its
dependence on a frozen normal mode, reads
\begin{equation}\label{eq_wire_k0}
\frac{\lambda_{a c}}{\Omega_\sigma^2 - \omega_a^2} + \frac{\lambda_{b c}}{\Omega_\sigma^2 - \omega_b^2}  = -1
\end{equation}
with $\Omega_\sigma$ frequency of the frozen mode (see details in SI).

This case is an example of the general result stating that
given any network, a part of it (in our case a linear motif,
$\mathcal{C}_1$) can be synchronized by tuning one of its
components, for instance a frequency or coupling of the motif.
A key point is that this is independent of the frequencies and
links of the rest of the network, provided the motif is
properly embedded in the network. The links between
$\mathcal{C}_1$ and the rest of the network should satisfy
\begin{equation}\label{eq_wire_coupl}
\left( \frac{\lambda_{a c}}{\Omega_s^2 - \omega_a^2} \right) \lambda_{a j} +
\left( \frac{\lambda_{b c}}{\Omega_s^2 - \omega_b^2} \right) \lambda_{b j}+
\lambda_{c j} = 0  \ ,\ \forall j \ .
\end{equation}
This is equivalent to saying that a synchronized motif with
robust quantum correlations can preserve these features when
linked to an $arbitrary$ network, if some constraints on the
reciprocal links are satisfied.  For instance, each node of
the synchronized motif needs to share with the rest of the
network more than one link. In Fig.
\ref{figure4} we compare the behavior of two linear motifs of
a large network, where a first motif $\mathcal{C}_1$ is
synchronized, satisfying
Eqs.\eqref{eq_wire_k0}-\eqref{eq_wire_coupl} while the second
one $\mathcal{C}_2$ is not. After a transient a frozen mode
tames the dynamics of  $\mathcal{C}_1$, which then shows a
synchronous evolution and robust correlations. It can also be
shown that quantum purity and energy reach higher values of
a stationary non-thermal state. This is
compared with the non-synchronized motif $\mathcal{C}_2$ whose
dynamics quickly relaxes to a thermal state. The case of a
three oscillator chain  is an example showing the possibility
to tune synchronization and quantum effects in a motif within
the network when the proper link conditions are satisfied.

\subsection{Entangling two oscillators through a network}

The same technique discussed in the previous section can be
applied to the case in which we aim to synchronize few, even if
not directly connected, elements of a network. But this does
not mean that any set of nodes can be synchronized
asymptotically. In fact, we find that for a CB we can only
synchronize two different and not directly linked
($\lambda_{ab}=0$) oscillators if we synchronize with them also
other intermediate linked elements (like in the linear motif
example, Fig. \ref{figure4}) or when these two oscillators are
identical, like we discuss in the present section.

\begin{figure}
\includegraphics[width=8cm]{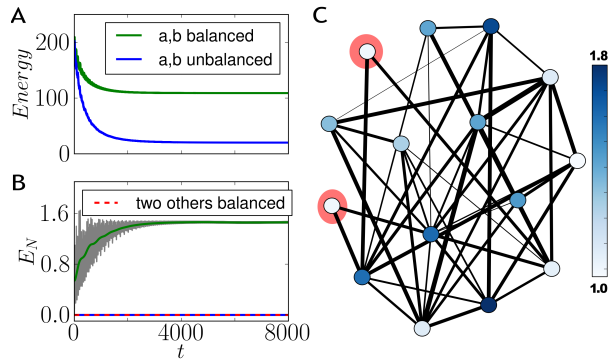}%
\caption{ (A) Energy evolution and (B) entanglement (logarithmic negativity) between two nodes with
identical frequency $\omega_0$ (we call these nodes $a$ and $b$ and are plotted
in red in the network displayed in panel (C)). The network is a random one (connection probability
$p=0.6$ of 15 oscillators and same frequency and couplings distribution as in Fig. \ref{figure4}. We 
compare the situations in which the couplings from the red nodes to the rest of the network (they are 
directly connected to other nodes called $c$ and $d$) are properly balanced in order to avoid dissipation
$(\lambda_{a c} = \lambda_{b c}= -0.15 \omega_0^2$ and $\lambda_{a d}  = \lambda_{b d}= -0.12 \omega_0^2)$
with the case when this balance is perturbed $(\lambda_{a c} + 0.04\omega_0^2$ and $\lambda_{a d} + 0.04
\omega_0^2)$ .The third line in panel (B) shows the entanglement between other two arbitrary oscillators
in the situation in which $a$ and $b$ are balanced.  Time units such that $\omega_0=1$ 
\label{figure5}}
\end{figure}

We consider the case of two identical oscillators (i.e. with
$\omega_a=\omega_b$) prepared in a separable state, with some
local squeezing. They are not directly coupled
($\lambda_{ab}=0$) but are connected through an arbitrary
network. In general they will dissipate  reaching the thermal
  state, but with the proper conditions we find an
important result: because their frequencies are identical it is
possible to construct a frozen normal mode involving only these
two nodes, given by $ Q_\sigma = \mathcal{F}_{a \sigma} ~q_a +
\mathcal{F}_{b \sigma} ~q_b,  $ with  $ \mathcal{F}_{a
\sigma},\mathcal{F}_{b \sigma} \neq 0$ and this can be
obtained, for instance, by properly attaching them to the
network. In other words, by tuning their couplings to the
network it is possible to have both oscillators  relaxing onto
a frozen mode, so that they will be synchronized and will keep
a higher energy than otherwise. Most importantly, in this case
entanglement can actually be generated between oscillators
initially in a separable state and remains high asymptotically.

In order to entangle the oscillators, their coupling to the rest of the network needs to
fulfill the  condition
\begin{equation}\label{eq_lobes}
\sum_{k=a,b} \mathcal{F}_{k \sigma} \lambda_{k j} = 0,
\end{equation}
(similar to Eq. \eqref{eq_wire_coupl}), achieved by proper
tuning of coupling strengths of the active links ($j$) with the
rest of the network $\lambda_{a j},\lambda_{b j}$. In Fig.
\ref{figure5}A and B we show the evolution of energy and
entanglement of  the oscillators $a$ and $b$ when linked to a
random network. As we see in  Fig. \ref{figure5}C there is not
direct link between $a$ and $b$ ($\lambda_{ab}=0$) and the
whole system dissipates in a common environment. The case where
the oscillators $a$ and $b$ are coupled to the network
following the prescription \eqref{eq_lobes} is compared to
another case in which their links are not properly balanced (we
slightly change the coupling strengths). Both
energy and entanglement are shown to be
 sensitive to the structure of the  reciprocal links and the
possibility to actually bring the added nodes into an entangled
state that will survive asymptotically is guaranteed by Eq.
\eqref{eq_lobes}.  The importance of this result is twofold: in
terms of applications it shows that it is possible to
dynamically generate entanglement between two non-linked nodes
embedded in a random network by tuning their connections to it,
and on the other hand it enlarges the scenario for asymptotic
entanglement generation through the environment. It is known
that large entanglement can be generated between far
oscillators during a transient due to a sudden-switch
\cite{Plenio} or through parametric driving \cite{Galve}. On
the other hand, a common environment leads to
entanglement between a pair of spins \cite{dfs}
or oscillators \cite{paz-roncaglia}.
%
%

\section{Discussion}

Our results on synchronization in dissipative harmonic networks and its
optimization give a flavor of all the possibilities that show up once the
mechanism behind the phenomenon is understood.  At difference from widely
considered self-sustained non-linear oscillators, here we focus in a linear
system showing how synchronization  can emerge after a transient for
dissipation processes introducing inhomogeneous decay rates among the systems
normal modes. A synchronous oscillation is predicted, for the first time,
either in a long transient during relaxation to the equilibrium state or in a
stationary non-thermal state.   We considered the most significant examples of
correlation length of the bath larger than the system size (CB) and of a node
of the network more strongly exposed to dissipation (LB), displaying
synchronous dynamics. On the other hand, for independent environments (SB) on
different nodes the resulting dynamics remains incoherent even when increasing
the strength of the reciprocal couplings in the network.

The presence of synchronization in the whole or a part of the network witnesses
the survival of quantum correlations and entanglement between the involved
nodes.  This connection between a coherent oscillation in the network and  its
non-classical state is a powerful result in the context of complex quantum
systems, considering the abundance of this  phenomenon. Indeed, the condition
underlying synchronization provides a strategy to protect a  system subspace
from decoherence. Our discussion and methodological approach are general, but
we show specific consequences of our analysis, such as global or partial
synchronization in a network through local tuning in one node (synchronizer) as
well as the possibility of connecting two nodes (not linked between them) to a
network and synchronize and entangle them, even starting from separable states.
Even if the reported results refer to random networks, our analysis applies to
generic ones, also including homogeneous and disordered lattices and do not
require all-to-all connectivity. 

In some sense, tuning part of a network so that the rest of it
reaches a synchronous, highly correlated state can be seen as a
kind of reservoir engineering, where here the tuned part of the
network would be a part of the reservoir. This is to be
compared with recent proposals of dissipative engineering for
quantum information, where special actions are performed
to target a desired non-classical state
\cite{Diehl,Barreiro2010,cirac2009,blatt2011}. 
In the context of quantum communications and considering recent results
on quantum Internet \cite{Kimble,Ritter},
our study can offer some insight
in designing a network with coherent information transport properties. 
Furthermore  implications of our approach can be explored in the context of
 efficient transport in biological systems
 \cite{Panitchayangkoon,Engel} .
An interesting methodological
connection is also with transport through (classical) random networks
\cite{transport}. On the other hand, our
analysis, when restricted to the classical limit, also gives
some insight about vibrations in an engineering context, 
providing the conditions for
undamped normal modes and their effect \cite{Rayleigh,Engineering}. 

%

\section{Methods}

\subsection{Interaction with the environment}

 The Hamiltonian of the system $H_S$ as defined in Eq. (\ref{HS})
 is diagonalized in the basis of its eigenmodes $\bf{Q}=\mathcal{F}^T\bf{q}$ yielding
 $\Omega=\mathcal{F}^T\mathcal{H}\mathcal{F}$.
 In a microscopic description with independent oscillators modeling the
environment, the system-bath interaction Hamiltonian for SB takes the form
\begin{equation}\label{HISB}
H_I^{SB} = -  \gamma \sum_{m=1}^{N}   q_m B^{(m)} \text{  , with }
B^{(m)}=\sum_{\alpha=1}^{\infty} \lambda_{\alpha} X_{\alpha}^{(m)},
\end{equation}
being  $\gamma$ the system-bath coupling strength
(explicitly shown  for the ease of understanding), 
$X_{\alpha}^{(m)}$ the position operators for each
environment oscillator $\alpha$ (representing for instance a
vibrational mode, or an optical one, etc...) of the bath $B^{(m)}$ in
which the network unit $(m)$ is dissipating.
 As explained in the main text, this situation occurs when the coherence length
 of the environment is smaller  than the spatial extension of the system. Thus
 each oscillator dissipates to its own heat bath. In the opposite case, a common bath
 is seen by all oscillators, resulting in an interaction Hamiltonian
 \begin{equation}\label{HICB}
  H_I^{CB} = -  \gamma \sum_{m=1}^{N}   q_m B,
\end{equation}
and actually involving only the average position (here the center
of mass) of the network, Fig. \ref{fig_1}B. Notice that in the eigenmodes basis

\begin{equation}\label{HICB_nm}
  H_I^{CB} =  - \gamma  \sum_m \kappa_m Q_m B \text{  , with }\kappa_m = \sum_n  \mathcal{F}_{n m}.
\end{equation} 
The \textit{effective couplings} $\kappa_m$ are different and determined by
characteristics of the network such as topology, coupling
strengths, and frequencies, as encoded in the diagonalization matrix $\mathcal{F}$. 
This is in stark contrast to the case of
identical SB \eqref{HISB} where all normal modes have equal effective
couplings to the baths (to see this, notice that we can transform the bath operators $X_\alpha^{(m)}$
to a new basis which exactly cancels the transformation $\mathcal{F}$; these new `oscillators' can be shown 
to have the same statistical properties as the others, thus resulting in equivalent heat baths).

Finally, the case of a given node $d$ dissipating much faster than any other is modeled by
\begin{equation}\label{HILB}
  H_I^{LB} = - \gamma q_d B.
\end{equation}
This local bath (LB) situation does also lead to non-uniform
environment interaction in some of the normal modes with
effective couplings ${\kappa_m}$:
\begin{equation}\label{HILB2}
  H_I^{LB} =  - \gamma \sum_m {\kappa_m} Q_m B \text{  , with }
  {\kappa_m} = \mathcal{F}_{d m}.
\end{equation}

All the mentioned situations of
separate, common and local bath can be described  by different
master equations \cite{Weiss} for the evolution of the network
state.

\subsection{Master equation}
A standard procedure allows to obtain from the total
Hamiltonian of system, bath and reciprocal interaction the
evolution of the reduced density matrix for the state of the
system, in our case a network of different oscillators. After a
(post-trace) rotating wave approximation, the master equations in the weak coupling limit 
for separate, common, and local baths are in the Lindblad form,
guarantying a well-behaved system dynamics. These
equations (given in SI) are obtained by generalization of the
problem of a pair of coupled oscillators \cite{PRAsync}. For
the purpose of our analysis it is interesting to consider the
master equation in the normal-mode basis $([Q_k,P_l]=i\delta_{k
l})$.
\begin{eqnarray}
&&\frac{d \rho(t)}{dt}=-i[H_S,{\rho}(t)]- \nonumber\\
&-& \frac{1}{4} \sum_{n} i \Gamma_{n} \left( [Q_n,\{P_n,\rho(t)\}] -[P_n,\{Q_n,\rho(t)\}]  \right) + \nonumber\\
 &+& D_{n}  \left( [Q_n,[Q_n,\rho(t)]] -  \frac{1}{\Omega_{n}^2} [P_n,[P_n,\rho(t)]] \right)
\end{eqnarray}
where $\Omega_n$ are the normal-mode frequencies of $H_S$ and
the damping and diffusion coefficients,  for an Ohmic bath 
with spectral density
$J(\omega) = (2\gamma/\pi)\omega  \Theta(\Lambda - \omega)$, 
are, at temperature $T$ far away from the frequency cutoff $\Lambda$:
$\Gamma_{n}= \kappa_n^2 \gamma$ and $D_{n} =  \kappa_n^2 \gamma
\Omega_n \coth(\frac{\Omega_n}{2 T}) $, being $\kappa_n$ the
effective couplings (see Eqs. \eqref{HICB_nm} and
\eqref{HILB2}). Boltzmann constant is taken to be unity. With
the appropriate definition of the couplings this equation is
valid both for common and local bath while for CB
we have: $\Gamma_{n}= \gamma$ and $D_{n} = \gamma \Omega_n
\coth(\frac{\Omega_n}{2 T}) $ i.e. we obtain the same damping
coefficient for all normal modes. The main differences in the
models of dissipation here proposed reside in these expressions
for the master-equation coefficients that will produce
different friction terms in the equations of motion determining
collective or individuals behaviors (See SI). We stress that
the choice of this master-equation representation is not
critical for our main conclusions, as also shown in
\cite{PRAsync}.

\subsection{Synchronization factor}
Synchronization between two time series $f(t)$ and $g(t)$ can
be characterized by a commonly used indicator, namely
$C_{f,g}(t,\Delta t)=\overline{ \delta f \delta
g}/\sqrt{\overline{\delta f^2} \ \overline{\delta g^2} }$ where
the bar stands for a time average $\overline{f}=\frac{1}{\Delta
t}\int_{t}^{t+\Delta t}dt'f(t')$ with time window $\Delta t$
and $\delta f=f-\overline{ f}$.  For `similar' and in phase
(anti-phase) evolutions  $C\sim 1$ ($-1$), while it tends to
vanish otherwise. As a figure of merit for global
synchronization in the whole network we look at the product
(neglecting the sign) of the indicator for all pairs of
oscillators in the system.
When the time series
correspond to positions second moments we have
 $\mathcal{S}=\Pi_{i<j}|C_{\langle
q_i^2\rangle\langle q_j^2\rangle}|$. This collective
synchronization factor $\mathcal{S}$ can reach unit value
only in presence of synchronous dynamics between all the pairs
of oscillators in the network.

\subsection{Mutual information, discord and entanglement}
 The total amount of correlations present in a bipartite system can be measured by the mutual
information $\mathcal{I}=S_A+S_B-S_{AB}$ with $S_i$ the entropy
of the reduced system $i=A,B$ and $S_{AB}$ the total entropy.
It can be decomposed in a purely classical part
$\mathcal{J}_{A:B}=\min [S_A-S_{A|E_j^B}]$, with
$S_{A|E_j^B}=\sum_ip_iS(\varrho_{A|E_i^B})$, $p_i={\rm
Tr}_{AB}(E_i^B\varrho)$ and $\varrho_{A|E_i^B}=
E_i^B\varrho/{p_i} $ 
(where $i$ labels the possible outcomes of a general measurement 
with operators $E_i^B$ acting on $B$ 
occurring with
probability $p_i$ and resulting in a density matrix $\varrho_{A|E_i^B}$), 
and a quantum part which is just the difference
$\delta_{A:B}=\mathcal{I}-\mathcal{J}_{A:B}$, the quantum
discord \cite{disc0,disc1,disc2}.
As a quantifier for entanglement we have
used the logarithmic negativity $E_N=\max(0,-\ln \nu_-)$, with
$\nu_-$ the smallest symplectic  eigenvalue of the partially
transposed density matrix \cite{horodecki_review}. 
 We quantify correlations between pairs of (linked or unlinked)
nodes by applying these definitions to different pairs of oscillators in the network.
Average
quantities ($\langle .\rangle $) for all pairs in the network can
be considered to assess globally the network.

\begin{acknowledgments} 
This work was partially supported by MINECO (Spain), FEDER, CSIC and Govern Balear
through projects FISICOS (FIS2007-60327), TIQS (FIS2011-23526) and  JdC and JAE programs.

\end{acknowledgments}

\section*{Author contributions}
RZ and FG planned the research. GM did most calculations and numerical simulations. All authors contributed analysing and discussing results and writing the manuscript. 

\section*{Additional Information}
Competing Financial Interests: None.

\end{document}